\documentclass[pra, showpacs, twocolumn]{revtex4}

\usepackage{amssymb, amsmath, amsthm}

\usepackage[dvipdfm]{hyperref}

\theoremstyle{remark}
\newtheorem{thm}{Theorem}

\newcommand{\C}{\mathbb{C}}
\newcommand{\N}{\mathbb{N}}

\DeclareMathOperator{\spn}{span}
\newcommand{\bra}[1]{\langle #1 |}

\newcommand{\ket}[1]{| #1 \rangle}
\newcommand{\dket}[1]{| #1 \rangle \! \rangle}
\newcommand{\bracket}[2]{\langle #1 | #2 \rangle}
\newcommand{\dbracket}[2]{\langle \! \langle #1 | #2 \rangle \! \rangle}

\begin{document}

\title{Solution to the Mean King's problem with mutually unbiased bases for arbitrary levels}
\author{Gen Kimura}
\email{gen@ims.is.tohoku.ac.jp}
\author{Hajime Tanaka}
\email{htanaka@ims.is.tohoku.ac.jp}
\author{Masanao Ozawa}
\email{ozawa@math.is.tohoku.ac.jp}
\affiliation{Graduate School of Information Sciences, Tohoku University, Aoba-ku, Sendai 980-8579, Japan}
\date{\today}

\begin{abstract}
The Mean King's problem with mutually unbiased bases is reconsidered for arbitrary $d$-level systems.
Hayashi, Horibe and Hashimoto [Phys. Rev. A \textbf{71}, 052331 (2005)] related the problem to the existence of a maximal set of $d-1$ mutually orthogonal Latin squares, in their restricted setting that allows only measurements of projection-valued measures.
However, we then cannot find a solution to the problem when e.g., $d=6$ or $d=10$.
In contrast to their result, we show that the King's problem always has a solution for arbitrary levels if we also allow positive operator-valued measures.
In constructing the solution, we use orthogonal arrays in combinatorial design theory.
\end{abstract}


\maketitle

\section{Introduction}

The \emph{Mean King's problem} is a problem to retrodict the outcome of a measurement of a basis randomly chosen from a maximal set of \emph{mutually unbiased bases} (MUBs) \cite{Schwinger,Ivanovic,WF}.
It was first introduced in \cite{VAA} for spin-$\frac{1}{2}$ systems, and later considered for systems with prime number levels \cite{AEEA} and prime power levels \cite{AD,HHH,KR}.
The problem is often stated as a tale \cite{AEEA,KR}:

``Once upon a time, there lived a mean King who loved cats.
The King hated physicists since the day when he first heard what had happened to Schr\"{o}dinger's cat.
One day, a terrible storm came on, and Alice, a physicist, got stranded on the island that was ruled by the King.
The King called Alice to the royal laboratory and gave her a challenge:
First, Alice can prepare a $d$-level quantum system (a $d$-level atom) in any state of her own liking and hand it over to the King.
The King will then secretly measure the atom with respect to one of $d+1$ mutually unbiased bases and return it to Alice.
Alice is then allowed to perform one more measurement on the atom.
Afterwards, the King reveals his measurement basis and then Alice must immediately guess the correct output of the King's measurement, or she will die a cruel
death.''

Here, the King's measurement is assumed to be a standard projective measurement of a basis $\{\ket{\varphi_m}\}_{m=0}^{d-1}$ of the atom system, so that the measurement in the state $\ket{\psi}$ leads to the output (index) $m$ with probability $|\bracket{\varphi_m}{\psi}|^2$ and leaves the system in the state $\ket{\varphi_m}$.
On the other hand, Alice is assumed to be allowed any measurement not restricted to that of a basis of the atom system.

The standard approach to the King's problem is to make use of entanglement \cite{VAA,AEEA,AD,KR,HHH}.
Alice prepares two $d$-level quantum systems $\C^d \otimes \C^d$, one to be handed over to the King and the other to be kept by Alice in secret, in a maximally entangled state.
After the King's measurement of one of $d+1$ MUBs, Alice is then supposed in the literature to carry out a measurement of projection-valued measure (PVM) on the space $\C^d \otimes \C^d$.

Under the above assumptions, Hayashi, Horibe and Hashimoto \cite{HHH} showed the equivalence of the existence of a solution to the King's problem and that of a maximal set of $d-1$ mutually orthogonal Latin squares, or equivalently, $d+1$ mutually unbiased striations \cite{Wootters}.
Then, it turns out that we cannot find a solution when e.g., $d = 6$ or $d = 10$, in which cases $d-1$ mutually orthogonal Latin squares do not exist, even if there might be a maximal set of $d+1$ MUBs;
the existence of the latter is still an open problem except for prime power levels.
The purpose of the present paper is to show that the King's problem always has a solution for arbitrary levels if we relax the above assumption to allow Alice to carry out any measurement of a positive operator-valued measures (POVM) on the same space $\C^d \otimes \C^d$.

The notion of POVM measurement was introduced by Helstrom \cite{Helstrom} to generalize conventional PVM measurements and to show that there is a class of optimization problems to which the optimum is achieved by a POVM measurement but not by any PVM measurements.
Nowadays, POVM measurement is considered as the most general description of measurement concerning the single measurement statistics, apart from the notion of instrument introduced by Davies and Lewis \cite{DL} that describes also the state change that determines the repeated or successive measurement statistics.
In virtue of the Naimark theorem, every POVM measurement can be realized by a PVM measurement of an extended system with the so-called ancilla;
see Holevo \cite{Holevo} for mathematical foundations of POVM measurements.
This is considered as a static realization with a non-local measurement.
A dynamical realization with a local measurement is obtained by the general realization theorem of completely positive instruments \cite{84QC}, so that any measurements can be realized as the unitary evolution of the composite system of the measured system and the probe followed by a subsequent PVM measurement of the probe.
Then, the difference between POVM measurements and PVM measurements arises only from the difference of the interaction or the probe preparation, and in some cases, a POVM measurement is more feasible than the corresponding PVM measurement, in particular, for measuring a continuous observable \cite{84QC} or for the measurement under conservation laws \cite{02CLU}.

As above, it is natural to assume that Alice can carry out, in principle, any POVM measurements on the same space $\C^d \otimes \C^d$ without considerable change of the resource allowed for her.
In this formulation, we first derive a simple criterion for the existence of a solution to the King's problem in Sec. \ref{sec:setting}.
Then in Sec. \ref{sec:construction}, we give a construction of a solution based on \emph{orthogonal arrays} in combinatorial design theory \cite{Colbourn}, instead of mutually orthogonal Latin squares.

We note, however, that our result gives no information on the existence of MUBs.
It is well-known that there can never be more than $d+1$ MUBs (cf. \cite{WF}).
There always exists a maximal set of $d+1$ MUBs when $d$ is a prime power \cite{Ivanovic,WF}, but a construction (or even the existence) of $d+1$ MUBs for other values of $d$ is a long-standing problem, even for the smallest case $d=6$.
For the rest of this paper, we just assume that we have a set of $k$ MUBs $\{\ket{A,a}_K\}_{a=0}^{d-1}$, $A\in\{0,1,\dots,k-1\}$, for the King's Hilbert space $\C^d$ (where $2\le k\le d+1$):
\begin{equation}\label{eq:MUBs}
	|\bracket{A,a}{A',a'}_K|^2=\delta_{A,A'}\delta_{a,a'}+(1-\delta_{A,A'})\frac{1}{d}.
\end{equation}
Of course, what we have in mind is the case $k=d+1$, but the problem itself makes sense even for smaller $k$ \cite{note:k=2}.

\section{Criterion for the Solution}\label{sec:setting}

We shall construct Alice's POVM on $\C^d\otimes\C^d$ from a suitable orthonormal basis $\{\ket{I}\}_{I=0}^{dd'-1}$ on a larger Hilbert space $\C^{d'}\otimes\C^d$ $(d'\ge d)$.
Let $V:\C^{d}\longrightarrow\C^{d'}$ be the natural isometric embedding of the space $\C^{d}$ into the extended space $\C^{d'}$.
Then, 
\begin{equation*}
	M_I\equiv (V\otimes \mathbb{I})^\dagger \ket{I}\bra{I}(V\otimes \mathbb{I}),
\end{equation*}
where $I\in\{0,1,\dots,dd'-1\}$, defines a POVM on $\C^d\otimes\C^d$.
(The exact value for $d'\in\N$ will be specified later.)

Following \cite{VAA,AEEA,AD,KR,HHH}, let Alice prepare the initial state in a maximally entangled state:
\begin{equation}\label{eq:initial_state}
	\ket{\Phi}\equiv \frac{1}{\sqrt{d}}\sum_{i=0}^{d-1}\ket{i}_A\otimes\ket{i}_K  \in\C^{d'}\otimes \C^d,
\end{equation}
with reference orthonormal bases $\{\ket{i}_A\}_{i=0}^{d'-1}$ and $\{\ket{i}_K\}_{i=0}^{d-1}$ for $\C^{d'}$ and $\C^d$, respectively.
Using any member $\{\ket{A,a }_K\}_{a=0}^{d-1}$ of the MUBs, \eqref{eq:initial_state} can be rewritten as
\begin{equation*}
	\ket{\Phi}=\frac{1}{\sqrt{d}}\sum_{a=0}^{d-1}\overline{\ket{A,a}}_A\otimes\ket{A,a}_K,
\end{equation*}
where $\overline{\ket{A,a}}_A\equiv\sum_{i=0}^{d-1}\bracket{i}{A,a}_K^* \ket{i}_A$.
If the King measured the basis $\{\ket{A,a}_K\}_{a=0}^{d-1}$ and obtained the output $a$, then the post measurement state will be
\begin{equation*}
	\ket{\Phi_{A,a}}\equiv\overline{\ket{A,a}}_A\otimes\ket{A,a}_K.
\end{equation*} 
We observe $\bracket{\Phi_{A,a}}{\Phi_{A',a'}}=\delta_{A,A'}\delta_{a,a'}+(1-\delta_{A,A'})/d$.

Here we remark the following.
Let $\omega\equiv\exp(2\pi i/d)$, and for $A\in\{0,1,\dots,k-1\}$ and $j\in\{0,1,\dots,d-1\}$ let
\begin{equation*}
	\ket{\widehat{\Phi}_{A,j}}\equiv\frac{1}{\sqrt{d}}\sum_{a=0}^{d-1}\omega^{aj}\ket{\Phi_{A,a}}.
\end{equation*}
Then $\ket{\widehat{\Phi}_{A,0}}=\ket{\Phi}$ and it is easy to see that
\begin{equation}\label{eq:orthogonality}
	\bracket{\widehat{\Phi}_{A,j}}{\widehat{\Phi}_{A',j'}}=\delta_{A,A'}\delta_{j,j'}+(1-\delta_{A,A'})\delta_{j,0}\delta_{j',0}.
\end{equation}
Let $\mathcal{A}$ be the one-dimensional subspace spanned by $\ket{\Phi}$, and for each $A\in\{0,1,\dots,k-1\}$ let $\mathcal{A}_A$ be the orthogonal complement of $\mathcal{A}$ in the linear span of $\ket{\Phi_{A,0}}$, $\ket{\Phi_{A,1}},\dots,\ket{\Phi_{A,d-1}}$, i.e., $\spn\{\ket{\Phi_{A,a}}\}_{a=0}^{d-1}=\mathcal{A}\oplus\mathcal{A}_A$.
Then, it follows from \eqref{eq:orthogonality} that $\mathcal{A}_A$ is spanned by $\ket{\widehat{\Phi}_{A,1}}$, $\ket{\widehat{\Phi}_{A,2}},\dots,\ket{\widehat{\Phi}_{A,d-1}}$, and Alice's space $\C^{d'}\otimes \C^d$ is decomposed into the orthogonal direct sum
\begin{equation}\label{eq:decomposition}
	\C^{d'}\otimes\C^d=\mathcal{A}\oplus\mathcal{A}_0\oplus\dots\oplus\mathcal{A}_{k-1}\oplus\mathcal{B},
\end{equation}
where
\begin{equation}\label{eq:B}
	\mathcal{B}\equiv\left(\spn\{\ket{\Phi_{A,a}}\}_{A=0,}^{k-1,}{}_{a=0}^{d-1}\right)^\bot.
\end{equation}
It seems that this structure is fundamental in the discussion of MUBs (cf. \cite{WF}).

With the above setting, now Alice has to find the basis $\{\ket{I}\}_{I=0}^{dd'-1}$ on $\C^{d'}\otimes\C^d$ and an \emph{estimation function} $s(I,A)\in\{0,1,\dots,d-1\}$, namely her guess for the King's output $a$ based on her output $I$ and the King's choice $A$.
For fixed $A$ and $a$, Alice's (conditional) success probability is then given by $\sum_{I=0}^{dd'-1}\delta_{a,s(I,A)}|\bracket{I}{\Phi_{A,a}}|^2$.
Thus, in order to save her life with certainty we must have \cite{HHH}
\begin{equation}\label{eq:condition1}
	\bracket{I}{\Phi_{A,a}}=0\quad\text{whenever}\quad s(I,A)\ne a.
\end{equation}
Now, we associate the basis $\{\ket{I}\}_{I=0}^{dd'-1}$ with a $dd'\times kd$ matrix $H$ defined by
\begin{equation*}
	H(I;A,a)\equiv\bracket{I}{\Phi_{A,a}}.
\end{equation*}
Then, obviously
\begin{equation}\label{eq:condition2}
	H(I;A,a)=0\quad\text{whenever}\quad s(I,A)\ne a,
\end{equation}
and moreover it follows that
\begin{equation}\label{eq:condition3}
	(H^{\dagger}H)(A,a;A',a')=\delta_{A,A'}\delta_{a,a'}+(1-\delta_{A,A'})\frac{1}{d}.
\end{equation}

Thus, we have shown that if we have an estimation function $s(I,A)$ and an orthonormal basis $\{\ket{I}\}_{I=0}^{dd'-1}$ for $\C^{d'}\otimes\C^d$ satisfying the survival condition \eqref{eq:condition1}, then there is a matrix $H$ such that \eqref{eq:condition2} and \eqref{eq:condition3} hold.
Now, we shall show the converse statement that given an estimation function $s(I,A)$ and a matrix $H$ satisfying \eqref{eq:condition2} and \eqref{eq:condition3}, we can find an orthonormal basis $\{\ket{I}\}_{I=0}^{dd'-1}$ for $\C^{d'}\otimes\C^d$ satisfying \eqref{eq:condition1}.
To show this, suppose that a function $s(I,A)\in\{0,1,\dots,d-1\}$ and a $dd'\times kd$ matrix $H$ satisfy \eqref{eq:condition2} and \eqref{eq:condition3}.
Let $\dket{\Psi_{A,a}}$ denote the $(A,a)$-th column vector of $H$.
Then, since $\dbracket{\Psi_{A,a}}{\Psi_{A',a'}}=\bracket{\Phi_{A,a}}{\Phi_{A',a'}}$, there is a unique unitary operator
\begin{equation*}
	U:\spn\{\ket{\Phi_{A,a}}\}_{A=0,}^{k-1,}{}_{a=0}^{d-1}\longrightarrow\spn\{\dket{\Psi_{A,a}}\}_{A=0,}^{k-1,}{}_{a=0}^{d-1}
\end{equation*}
such that
\begin{equation}\label{eq:isometry}
	U\ket{\Phi_{A,a}}=\dket{\Psi_{A,a}}.
\end{equation}
Specifically, $U$ is determined by $U\ket{\widehat{\Phi}_{A,j}}\equiv\dket{\widehat{\Psi}_{A,j}}$, where
\begin{equation*}
	\dket{\widehat{\Psi}_{A,j}}\equiv\frac{1}{\sqrt{d}}\sum_{a=0}^{d-1}\omega^{aj}\dket{\Psi_{A,a}}.
\end{equation*}
Now, arbitrarily extend $U$ to a unitary operator
\begin{equation}\label{eq:extended_operator}
	\tilde{U}:\C^{d'}\otimes\C^d\longrightarrow\C^{dd'},
\end{equation}
and let
\begin{equation*}
	\ket{I}\equiv\tilde{U}^\dagger\dket{I},
\end{equation*}
where $\{\dket{I}\}_{I=0}^{dd'-1}$ denotes the standard basis for the column space $\C^{dd'}$.
Then $\{\ket{I}\}_{I=0}^{dd'-1}$ is an orthonormal basis for $\C^{d'}\otimes\C^d$ and by \eqref{eq:isometry} we have $\bracket{I}{\Phi_{A,a}}=\dbracket{I}{\Psi_{A,a}}=H(I;A,a)$ and thus \eqref{eq:condition1} holds.

To summarize, we have the following.
\begin{thm}\label{thm:criterion}
Given an estimation function $s(I,A)\in\{0,1,\dots,d-1\}$, there exists an orthonormal basis $\{\ket{I}\}_{I=0}^{dd'-1}$ for $\C^{d'}\otimes\C^d$ satisfying \eqref{eq:condition1} if and only if there is a $dd'\times kd$ matrix $H$ such that \eqref{eq:condition2} and \eqref{eq:condition3} hold.
\end{thm}

\section{Orthogonal Arrays and the Existence of a Solution}\label{sec:construction}

An \emph{orthogonal array} of \emph{degree} $k$, \emph{order} $d$ and \emph{index} $n$, denoted $\mathrm{OA}_n(k,d)$, is an $nd^2\times k$ array with entries from $\{0,1,\dots,d-1\}$ such that every pair of symbols from $\{0,1,\dots,d-1\}$ occurs exactly $n$ times as a $1\times 2$ submatrix in the $nd^2\times 2$ matrix consisting of any pair of two distinct columns chosen from the array (cf. \cite{Colbourn}).
It follows immediately from the definition that every symbol occurs exactly $nd$ times in each column of an $\mathrm{OA}_n(k,d)$.
Thus, in other words, an $nd^2\times k$ array $T$ with entries from $\{0,1,\dots,d-1\}$ is an $\mathrm{OA}_n(k,d)$ if and only if
\begin{equation}\label{eq:OA}
	\frac{1}{nd}\sum_{I=0}^{nd^2-1}\delta_{a,T(I,A)}\delta_{a',T(I,A')}
=\delta_{A,A'}\delta_{a,a'}+(1-\delta_{A,A'})\frac{1}{d}.
\end{equation}
(Compare this equation with \eqref{eq:MUBs}.)
We note that an $\mathrm{OA}_1(k,d)$ is equivalent to a set of $k-2$ mutually orthogonal Latin squares of side $d$.
An $\mathrm{OA}_1(4,3)$ is given in Fig. \ref{fig:oa}.
\begin{figure}
\ttfamily
\begin{tabular}{|c|c|c|c|}
	\hline
	0 & 0 & 0 & 0 \\ \hline
	0 & 1 & 1 & 1 \\ \hline
	0 & 2 & 2 & 2 \\ \hline
	1 & 0 & 1 & 2 \\ \hline
	1 & 1 & 2 & 0 \\ \hline
	1 & 2 & 0 & 1 \\ \hline
	2 & 0 & 2 & 1 \\ \hline
	2 & 1 & 0 & 2 \\ \hline
	2 & 2 & 1 & 0 \\ \hline
\end{tabular}
\caption{\label{fig:oa} An $\mathrm{OA}_1(4,3)$.
We may think of the first two columns as representing the row and column indices of $3\times 3$ matrices, respectively.
Then the third and fourth columns correspond to two mutually orthogonal Latin squares of side $3$.
}
\end{figure}
Hayashi et al. \cite{HHH} constructed an estimation function using a maximal set of $d-1$ mutually orthogonal Latin squares.
There always exist $d-1$ mutually orthogonal Latin squares if $d$ is a power of a prime, but as mentioned in the introduction, it is known that this is not the case for some other values of $d$, such as $d=6$ and $d=10$.
On the other hand, we can always find an orthogonal array for each $k$ and $d$.
In fact, the array obtained by arranging all the $k$-tuples (in e.g. the lexicographic order) obviously defines an $\mathrm{OA}_{d^{k-2}}(k,d)$.
In particular, it is known \cite{Hanani} that there exists an $\mathrm{OA}_n(7,6)$ for all $n\ge 2$ \cite{note:OA}.

Now, we construct a solution to the King's problem based on orthogonal arrays.
Set $d'=nd$ and let $\left[s(I,A)\right]_{I=0,}^{nd^2-1,}{}_{A=0}^{k-1}$ form an $\mathrm{OA}_n(k,d)$.
We define an $nd^2\times kd$ matrix $H$ by
\begin{equation*}
	H(I;A,a)\equiv\frac{1}{\sqrt{nd}}\delta_{a,s(I,A)}.
\end{equation*}
Then by \eqref{eq:OA} $H$ satisfies \eqref{eq:condition2} and \eqref{eq:condition3}, and Theorem \ref{thm:criterion} shows that there is a solution to the problem.
In fact, the proof of Theorem \ref{thm:criterion} yields a somewhat explicit formula for the corresponding basis $\{\ket{I}\}_{I=0}^{nd^2-1}$ in this case.
Let $\{\ket{\Xi_b}\}_{b=0}^{e-1}$ be an orthonormal basis for $\mathcal{B}$, where $e\equiv\dim\mathcal{B}=nd^2-k(d-1)-1$.
Then by \eqref{eq:decomposition} we have
\begin{equation*}
	\ket{I}=\bracket{\Phi}{I}\ket{\Phi}+\sum_{A=0}^{k-1}\sum_{j=1}^{d-1}\bracket{\widehat{\Phi}_{A,j}}{I}\ket{\widehat{\Phi}_{A,j}}+\sum_{b=0}^{e-1}\bracket{\Xi_b}{I}\ket{\Xi_b}.
\end{equation*}
Since $\ket{\widehat{\Phi}_{A,0}}=\ket{\Phi}$, $\bracket{\Phi}{I}=1/(d\sqrt{n})$ and 
\begin{align*}
	\sum_{j=0}^{d-1}\bracket{\widehat{\Phi}_{A,j}}{I}\ket{\widehat{\Phi}_{A,j}}&=\frac{1}{d\sqrt{nd}}\sum_{j=0}^{d-1}\sum_{a=0}^{d-1}\omega^{(a-s(I,A))j}\ket{\Phi_{A,a}} \\
	&= \frac{1}{\sqrt{nd}}\ket{\Phi_{A,s(I,A)}},
\end{align*}
we find
\begin{equation}\label{eq:formula}
	\ket{I}=\frac{1}{\sqrt{n}}\ket{I'}+\sum_{b=0}^{e-1}\bracket{\Xi_b}{I}\ket{\Xi_b},
\end{equation}
where
\begin{equation*}
	\ket{I'}\equiv\frac{1}{\sqrt{d}}\sum_{A=0}^{k-1}\ket{\Phi_{A,s(I,A)}}-\frac{k-1}{d}\ket{\Phi}.
\end{equation*}
(Compare this expression with Eq. (10) in \cite{HHH}.)

\section{Example}

We illustrate the above construction of a solution to the problem in the case of the (trivial) $\mathrm{OA}_{d^{k-2}}(k,d)$.
Each $I\in\{0,1,\dots,d^k-1\}$ has a unique $d$-adic expansion:
\begin{equation*}
	I=\sum_{A=0}^{k-1}I_Ad^A\quad\text{where}\quad I_A\in\{0,1,\dots,d-1\}.
\end{equation*}
We define the array $\left[s(I,A)\right]_{I=0,}^{d^k-1,}{}_{A=0}^{k-1}$ by $s(I,A)\equiv I_A$.

In order to carry out the construction \eqref{eq:formula} of the basis $\{\ket{I}\}_{I=0}^{d^k-1}$ explicitly, we must specify $\tilde{U}\ket{\Xi_b}$ for an orthonormal basis $\{\ket{\Xi_b}\}_{b=0}^{e-1}$ for $\mathcal{B}$ in \eqref{eq:B} (where $e=d^k-k(d-1)-1$).
The space $\mathcal{B}$ depends on the particular set of MUBs.
When $k=d+1$ for instance, $\mathcal{B}$ is spanned by $\{ \ket{i}_A \otimes \ket{j}_K\}_{i=d,}^{d^d-1,}{}_{j=0}^{d-1}$.

For each $J\in\{0,1,\dots,d^k-1\}$ let
\begin{equation*}
	\dket{\widehat{\Psi}_J}\equiv\frac{1}{\sqrt{d^k}}\sum_{I=0}^{d^k-1}\omega^{\sum_{A=0}^{k-1}I_AJ_A}\dket{I}.
\end{equation*}
Then $\{\dket{\widehat{\Psi}_J}\}_{J=0}^{d^k-1}$ forms an orthonormal basis for the column space $\C^{d^k}$.
Note that $\dket{\widehat{\Psi}_{jd^A}}=\dket{\widehat{\Psi}_{A,j}}\quad\text{for}\quad j \in\{0,1,\dots,d-1\}$.
Let $\Theta:\{0,1,\dots,e-1\}\longrightarrow\{J:|\{A:J_A\ne 0\}|\ge 2\}$ be any bijection and define the extension $\tilde{U}:\C^{d^{k-1}}\otimes\C^d\longrightarrow\C^{d^k}$ of $U$ in \eqref{eq:extended_operator} by setting $\tilde{U}\ket{\Xi_b}\equiv\dket{\widehat{\Psi}_{\Theta(b)}}\quad\text{for}\quad b\in\{0,1,\dots,e-1\}$.
Then we find
\begin{equation*}
	\ket{I}=\frac{1}{\sqrt{d^{k-2}}}\ket{I'}
+\frac{1}{\sqrt{d^k}}\sum_{b=0}^{e-1}
\omega^{-\sum_{A=0}^{k-1}I_A\Theta(b)_A}\ket{\Xi_b}.
\end{equation*}

\section{Concluding remarks}
In contrast to the results in \cite{HHH}, we showed that for any $d$ we can always find a solution to the King's problem by performing a suitable POVM measurement, instead of a PVM measurement.
We note that our method in this paper also indicates how Alice constructs that POVM:
She just prepares a $d'(=nd)$ level ancilla to maximally entangle the $d$-level atom, and carries out the PVM measurement with respect to $\{\ket{I}\}_{I=0}^{dd'-1}$ constructed in the previous sections based on orthogonal arrays.

The authors are grateful to Akihiro Munemasa and Masahiro Hotta for fruitful discussions and comments on the Mean King's problem and the structure of MUBs.
GK and HT are JSPS Research Fellows.
MO is supported in part by the SCOPE project of MPHPT of Japan.


\begin{thebibliography}{10}
\bibitem{note:k=2}
We remark that, in the case $k=2$, Alice can in fact survive with probability $1$ without any ancilla at the stage of state preparation.
For instance, take any one of the pure states in $\{\ket{0,a}_K\}_{a=0}^{d-1}$ as the initial state and perform the PVM with respect to $\{ \ket{1,a}_K\}_{a=0}^{d-1}$.
\end{thebibliography}
\end{document}